\documentclass[showpacs,prb,floatfix,superscriptaddress,twocolumn,amsmath]{revtex4} 
\usepackage[T1]{fontenc}
\usepackage[latin1]{inputenc}
\usepackage[english]{babel}
\usepackage{epsfig}
\usepackage{amssymb}  %   MATH
\usepackage{amsmath}  %   MATH
\usepackage{latexsym} %   MATH
\usepackage{amsthm}   %   MATH

\begin{document}\title{Nematic phase without Heisenberg physics in
  FeAs planes}        
%\begin{document}\title{Nematic phase induced by finite size effects in  FeAs planes}       

\author{M. Capati}
\affiliation{Dipartimento di Fisica, Università di Roma ``Sapienza'', 
Piazzale Aldo Moro 2, I-00185 Roma, Italy}
\author{M. Grilli}
\affiliation{Dipartimento di Fisica, Università di Roma ``Sapienza'', 
Piazzale Aldo Moro 2, I-00185 Roma, Italy}
\affiliation{Istituto dei Sistemi Complessi, CNR, Dipartimento di Fisica, 
Università di Roma ``Sapienza'', Piazzale Aldo Moro 2, I-00185 Roma, Italy}
\affiliation{CNISM, Unit\'a di Roma ``Sapienza'', 
Piazzale Aldo Moro 2, I-00185 Roma, Italy}
\author{J. Lorenzana}
\affiliation{Dipartimento di Fisica, Università di Roma ``Sapienza'', 
Piazzale Aldo Moro 2, I-00185 Roma, Italy}
\affiliation{Istituto dei Sistemi Complessi, CNR, Dipartimento di Fisica, 
Università di Roma ``Sapienza'', Piazzale Aldo Moro 2, I-00185 Roma, Italy}

\begin{abstract}
We use Monte Carlo simulations and analytical arguments to 
analyze a frustrated Ising model with nearest neighbour
antiferromagnetic coupling $J_1$ and next nearest neighbour coupling
$J_2$. The model is  inspired on the physics of pnictide
superconductors and to some extent we argue that it can be more
representative of this physics than the Heisenberg counterpart. 
Parameters are chosen such that the ground
state is a columnar or striped state, as observed experimentally,  but
is close to the transition to
the simple N\'eel ordered antiferromagnetic state $R = J_2/|J_1|\gtrsim 0.5$.
We find that a nematic phase is induced close to $R = 0.5$ by finite
size effects and argue that this explains experiments in imperfect
samples which find a more robust nematic state as the quality of the
sample decreases [A. Jesche {\it et al.}, Phys. Rev. B {\bf 81},
134525 (2010)].
Including the effect of a weak coupling with the lattice we find that 
a structural transition occurs associated with a nematic phase, with a 
magnetic transition occurring at a lower temperature. These two transitions
merge into a single structural and
 magnetic transition with a stronger first-order character for 
larger spin-lattice couplings. These two situations are  in agreement
with the different phenomenologies found in different families of
pnictides. 
\end{abstract}

\date{\today} 
%\pacs{75.10.Hk Classical spin models; 75.10.-b General theory and models of magnetic ordering;
% 75.40.Cx Static properties (order parameter, static susceptibility, 
% heat capacities, critical exponents, etc.); 74.70.Xa 	Pnictides and chalcogenides;
% 75.40.Mg Numerical simulation studies }
\maketitle

\newcommand{\ie}{\begin{equation}}
\newcommand{\fe}{\end{equation}}
\newcommand{\iea}{\begin{eqnarray}}
\newcommand{\fea}{\end{eqnarray}}
\newcommand{\ia}{\begin{array}}
\newcommand{\fa}{\end{array}}

%Greek Letters

\def\a{\alpha}
\def\b{\beta}
\def\c{\xi}
\def\e{\varepsilon}
\def\d{\delta}
\def\g{\gamma}
\def\m{\mu}
\def\l{\lambda}
\def\L{\Lambda}
\def\f{\phi}
\def\n{\nu}
\def\o{\omega}
\def\p{\pi}
\def\s{\sigma}
\def\G{\Gamma}
\def\D{\Delta}
\def\O{\Omega}
\def\ra{\rightarrow}
\def\up{\uparrow}
\def\pll{\parallel}
\def\down{\downarrow}
\def\ran{\rangle}
\def\lan{\langle}
\def\Ra{\Rightarrow}
\def\pd{\partial}
\def\bk{{\bf k}}
\def\nn{\nonumber}
\def\ol{\overline}

\section{Introduction}

The FeAs high-Tc materials \cite{kamihara,rotter} are a very interesting new
playground to study  the 
interplay between lattice, magnetism and superconductivity. 
Quite generically undoped or slightly doped samples show a magnetic 
 phase which breaks the $C_4$ symmetry of the lattice,
consisting of magnetic moments aligned  ferromagnetically on one
direction and antiferromagnetically in the other direction.
This  ``columnar'' or ``spin-stripe'' phase is
accompanied by an orthorhombic distortion of the high
temperature tetragonal lattice. 

From symmetry considerations it is quite natural to assume that the 
structural transition from the high temperature tetragonal phase to
the orthorhombic phase is driven by the magnetism. However `1111'
materials like LaOFeAs\cite{delaCruz2008Magnetic} display 
the structural  transition at a higher temperature than the magnetic
transition.  This has led some authors to propose that the structural
transition drives the magnetism. 
Even more surprising is the fact that the temperature splitting between  
structural and magnetic transitions decreases upon increasing the sample 
quality\cite{Jesche2010Coupling} and the transport anisotropy is
reduced.\cite{liang} 

The stripe-magnetic phase, breaking both the $C_4$ symmetry and the
translation symmetry can be called ``smectic'',  in analogy with
the crystal phase that elongated molecules form in liquid crystals.
However in the last few years the question has emerged whether a ``nematic''
phase can also occur in these systems. The term ``nematic'', which is also 
borrowed from the field of liquid crystals, indicates a phase where
the rotational 
symmetry is broken, while the translational symmetry is fully preserved.
In the pnictide case, for instance, the square lattice $C_4$ 
symmetry of the FeAs layers in the tetragonal phase could be reduced to $C_2$
by the occurrence of a purely magnetic nematic phase. In turn this nematic state
could involve an interplay between structural and magnetic properties giving 
rise to structural and lattice signatures even in the absence of a ``smectic''
magnetic order. This issue not only has been addressed from the 
theoretical point of view 
\cite{Fang2008Theory,xusachdev,Vojta2009Lattice,fernandes2},
but increasing evidence is now experimentally emerging of a nematic state
above the magnetic transition in pnictides. Scanning tunnelling microscopy studies
detect quasiparticle electronic states having a reduced $C_2$ 
symmetry \cite{zhou} as well as local anisotropies \cite{chuang}, 
transport experiments in detwinned 122 crystals find an in-plane anisotropy, 
which cannot be attributed to lattice distortions only
\cite{chu,tanatar}, while optical experiments
find anisotropic mid-infrared spectra \cite{dusza}. If such nematic phase
exists, the coupling between the magnetic nematic state and the lattice 
would naturally account for the occurrence in some
materials (like 1111 or Ba(Fe$_{1-x}$Co$_x$)$_2$As$_2$)
of the tetragonal-orthorhombic transition at a higher temperature than the 
magnetic transition. The coupling between nematicity and lattice 
has also been taken as a
possible explanation for changes in elastic properties observed by 
ultrasound spectroscopy in FeAs systems \cite{fernandes1}, and the suppression 
of  orthorombicity in the superconducting phase \cite{nandi}.

An important question is if an electronic nematic phase drives the 
lattice distortion or vice versa. 
The split transitions occur only when the thermal transition is
continuous or very close to continuous. On the other hand most `122'
materials like SrFe$_2$As$_2$\cite{Krellner2008Magnetic} 
show a first-order transition in which simultaneously the structural
and the magnetic order parameter become non-zero. 

The more straightforward explanation for nematic phases starts from
 a two-dimensional  frustrated Heisenberg model\cite{Fang2008Theory}
 and is based on 
 theoretical results that dates back two decades.\cite{larkin}
This explanation relays on the order by disorder mechanism by which the
degeneracy of the frustrated Heisenberg model at the classical level 
is broken by thermal and quantum fluctuations. As a consequence a
nematic phase is 
stabilized which has Ising symmetry.  In two dimensions
antiferromagnetic (AF) long-range order  can only occur at zero temperature
while the nematic phase, belonging to the Ising universality class,
can occur at finite temperatures thus a large
region of temperatures exists where the system is in a nematic
phase. Interlayer coupling stabilizes a three dimensional AF state but
always with a magnetic transition occurring at a lower temperature
than the Ising nematic transition.\cite{Fang2008Theory}

While this explanation is very appealing it is worth to examine
whether the Heisenberg physics is really necessary and if there can be
other mechanism by which nematic phases can be stabilized. One
motivation to do so is that the region where the Heisenberg physics
is relevant can be quite narrow around the N\'eel temperature. 
Indeed stoichiometric FeAs materials have small magnetic moments and
are metallic which indicates rather itinerant magnetism\cite{Zhang2010Itinerant}. 
The magnetism thus is rather collective which means that a 
given site does not interact only with the
neighbors but with several sites over a coherence distance  
$\xi_0\gg a$, with $a$ the lattice spacing.  To observe 
Heisenberg-like critical
fluctuations the magnetic correlation length $\xi$ has to exceed 
$\xi_0$ which occurs only very close to the magnetic
critical temperature. If $\xi_0$ is
sufficiently large a small Ising like anisotropy or a three
dimensional coupling will make the 2D Heisenberg  
physics and the Goldstone modes  rather irrelevant. One way to
suppress from the start the presence of the Goldstone modes and have a
finite critical temperature is to consider Ising spins.

In this work we study a frustrated two-dimensional Ising model.
For simplicity we assume localized spins  which interact through effective
nearest-neighbor and next-nearest-neighbor antiferromagnetic
interactions  ($J_1>0$ and $J_2>0$ respectively) (see,
e.g. Ref. \onlinecite{Yildirim2008Origin}).   We thus obtain a frustrated 
Ising model without any interaction between the spins and the lattice and only at a 
later stage we will introduce a spin-lattice coupling. The model's
phase diagram is studied using Monte Carlo simulations and finite size
analysis. 

This model is interesting in its own and although its magnetic properties have already 
been extensively investigated\cite{Lopez1993Firstorder}, our analysis
will evidentiate 
new features and it will find interesting connections with the real
pnictide systems. We show that the critical behavior corresponds
closely to a 4-component Potts model.

For $0<R=J_2/|J_1|<0.5$,  
it is well known that there is a second-order transition toward 
a low temperature N\'eel antiferromagnetic phase. For $R>0.5$ the
system shows a magnetic transition toward the ``spin-stripe'' phase
observed in FeAs planes and mentioned above. 
 While in other cases, like in manganites, this antiferromagnetic spin
 configuration 
has been named c-type antiferromagnetism, we will keep the ``spin-stripe'' phase
nomenclature as usual in the pnictide field.

Our analysis aims to explore
the possibility of having nematic phases when all the 
Heisenberg physics is suppressed.
Although the model is very simplified we show below that it 
explains well several aspect of the phenomenology of FeAs
planes. In particular we find that nematic phases can be stabilized at
finite temperatures by finite size effects. We argue that such effect
can explain the larger tendency of ``bad samples'' to display nematic
phases. 

The paper is structured as follows. In Sect. II we introduce the model
and the technical framework to solve it. In Sect. III we report the 
results in the absence of the spin-lattice coupling, which is instead
introduced in Sect. IV. Our concluding remarks are in Sect. V.

%We will then take into account the lattice deformations occurring in real pnictide materials 
%introducing a spin-lattice coupling in our simplified model. We will
%see that a nematic  
%state above the magnetic state is strictly related to a structural transition at a
%higher temperature than the magnetic one. Indeed a nematic state implies the non equivalence 
%of the two principal lattice directions as it occurs in the orthorhombic lattice 
%(rectangular symmetry in a two dimensional lattice). Therefore, the occurrence of a nematic 
%phase can be related to the temperature splitting between the structural and magnetic 
%transitions observed in some of the iron-pnictide materials.

\section{Model and methods}

We consider a two-dimensional frustrated Ising model with
antiferromagnetic nearest-neighbour  (nn) and  next-nearest-neighbor
(nnn) interactions. The Hamiltonian of the
model reads
\iea\label{eq:H_frustrates_ising}
H_M(\{\s_i\}) &=& \sum_{i=1}^N \sum_{\d = x,y} J_1 \, \s_i \s_{i + \d} \nn \\
&& + \sum_{i=1}^N \sum_{\eta = x+y, -x+y} J_2 \, \s_i \s_{i+\eta},
\fea
with $J_1$ and $J_2$ both positive. The notation $i + x$ ($i+x+y$) indicates 
the first (second) neighbour of site $i$ in the $x$ ($x+y$)
direction. One can use a canonical transformation $\s_i \rightarrow
(-1)^{x_i + y_i} \s_i$ to change the sign of the first term. This
means that since the model is classical it is fully equivalent to 
a nn ferromagnetic model with AF frustration among nnn which has been
extensively studied in the literature\cite{Lopez1993Firstorder}. 
%With this transformation we must consider $R= J_2/|J_1|$, to have positive values of $R$.  

\begin{figure}[htbp]
\centering
 \epsfig{file=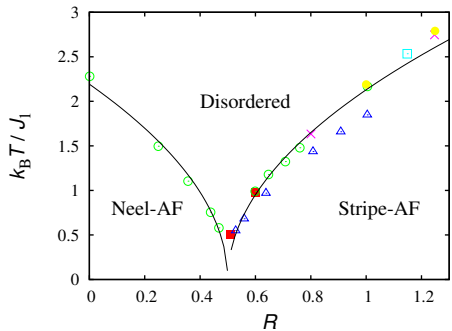,width=6.5cm}
\caption{(Color online) Phase diagram of the system. Each point represents the critical 
temperature for a given value of $R$, and the types of point depends on the different 
methods used to compute them. Empty circles (green online) have been obtained with Monte Carlo;
empty triangles (blue online) have been obtained with real space renormalization group; crosses (purpleonline)
indicate have been obtained with Monte Carlo renormalization group; the solid (yellow online) circles have been obtained
with series expansion.
The open square (light blue online) marks 
the tricritical point at $R=1.144$\cite{Lopez1993Firstorder}. 
The solid square (red online) mark the critical temperature values obtained with our 
MC simulations. The model studied in Ref.~\onlinecite{Lopez1993Firstorder} considers 
ferromagnetic nn interactions so that the ordered phase up to $R=0.5$ should 
more precisely be named ``N\'eel-F''.}
\label{fig:phase_diagram}
\end{figure}

If frustration is strong enough ($R > 0.5$), 
the ground state is given by a spin-stripe configuration. This configuration breaks 
both rotational symmetry and traslational symmetry in one direction. This state is 
characterized by a staggered magnetization with $(\p,0)$ or $(0,\p)$ wave vectors, 
which, for one configuration, is defined by
\begin{subequations}
\ie\label{eq:mpi0}
m_{(\p,0)} = \frac{1}{N} \sum_{j} \s_j \, {(-1)}^{x_j};
\fe
\ie\label{eq:m0pi}
m_{(0,\p)} = \frac{1}{N} \sum_{j} \s_j \, {(-1)}^{y_j},
\fe
\end{subequations}
We notice that at $T=0$, for a given orientation of the spin stripes, only one of the two staggered 
magnetizations is non zero. The sign of the two parameters depends on the position of the 
spin up stripes (odd or even rows or columns). In Fig.\ref{fig:phase_diagram} we report the phase 
diagram of the model in Eq.(\ref{eq:H_frustrates_ising}) as obtained
by Mor\'an-L\'opez et al.\cite{Lopez1993Firstorder} studying the equivalent nn ferromagnetic version. 
Here we have added two points 
obtained from our MC calculations.
 More detailed data including
nematic and lattice effects will be shown below.  
In Ref. \onlinecite{Lopez1993Firstorder} it was found that the transition toward the 
spin-stripe phase is weakly first-order up to a tricritical point at $R=1.144$. For 
higher values of $R$ the transition becomes of the second-order.

%We will search for nematic states at finite temperatures.  This can be
%though of as a  state in which thermally fluctuating  domains with stripe configuration, 
%contributing with opposite signs to macroscopic staggered magnetization. This latter  thus
%stays vanishingly small. This state breaks $C_4$ rotational symmetry only, while
%traslational symmetry is preserved in directions. 

\begin{figure}[htbp]
\centering
 \epsfig{file=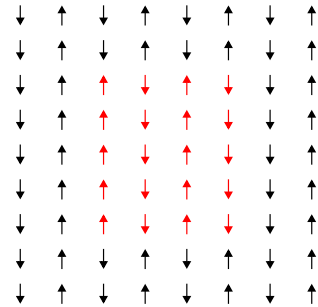,width=4.0cm,height=4.0cm}
\caption{(Color online) 
Illustration of a spin domain (gray arrows, red online), in 
which the staggered magnetization in {\it x} direction has opposite sign
with respect to the embedding cluster (black arrows).}
\label{fig:goccia_nematica}
\end{figure}

In Fig.~\ref{fig:goccia_nematica} we show a domain of the perfect
stripe order in which the directionality is preserved but the phase of
the AF changes by $\pi$. At finite temperatures it can happen that 
domains of this kind are  preferentially excited destroying the long-range magnetic order  
but still breaking $C_4$ rotational symmetry so that the system
reaches a nematic state.  

Following Ref.~\onlinecite{Vojta2009Lattice} we define a nematic
magnetization which measures the global breaking of $C_4$ symmetry and
 which for one configuration reads, 
\ie\label{eq:mphi}
m_{\phi} = \frac{\sum_i \phi_i}{N}.
\fe
Here the local nematic parameter
$\phi_i$ is defined by
\ie
\phi_i = \frac{1}{2} \s_i (\s_{i+x} - \s_{i+y}).
\fe
The magnetic and nematic order parameters are defined as the thermal
averages $\langle ... \rangle$
of Eqs.~(\ref{eq:mpi0}),(\ref{eq:m0pi}),(\ref{eq:mphi}). The range of
all these three parameters is $[-1:1]$.  

The ``nematic phase'' is 
characterized by finite values of the nematic magnetization, consequence of the 
break of the rotational symmetry $C_4$, but vanishing values of the
staggered magnetization. It is customary to call $\langle m_\f \rangle$ the nematic order parameter\cite{Vojta2009Lattice}
and we follow this convention here, although it is different from zero
also in the spin-stripe or smectic phase which, strictly  speaking, are
not nematic phases.  The sign of  
$\langle m_\f \rangle$ depends on the preferred orientation of the spin stripes.  
Table~\ref{tableI} summarizes the above discussion of the order
parameters.

%Our purpose is to check the
%possible existence of a nematic phase in a finite range of temperatures and to 
%compute the critical indexes of nematic (if any) and magnetic transitions.  
%The following table summarizes the above discussion of the order parameters
\begin{table}
\begin{center}
\begin{tabular}{l| c| c}%{||p{1cm}||*{3}{c|}|}
\hline
                  &   $\langle m_{(\p,0)}\rangle$, $\langle m_{(0,\p)}\rangle$   &
                  $\langle m_\f\rangle$     \\
\hline
\hline
Disorder          &        0        &      0     \\
\hline
Magnetic phase    &        $\neq 0$          &      $\neq 0$        \\
\hline
Nematic phase     &        0        &      $\neq 0$        \\
\hline
\end{tabular}
\end{center}
\label{tableI}
\end{table}

To numerically implement our analysis we define a square matrix (that is our lattice), 
every element of which may assume the values $\pm 1$, according to the up or down orientation of
the spin on the site. Physical quantities like staggered magnetizations, 
nematic order parameter, spatial correlation functions and susceptibilities, are 
calculated using the MC method\cite{Newman1999Monte,Katzgraber2010Introduction}. This 
method exploits the Metropolis algorithm to generate, in a random way, a chain of states 
(called ``Markov chain''), which are distributed according to the Boltzmann distribution 
function. Thus the MC procedure is able to simulate the thermal fluctuations of the physical 
quantities upon exploring the phase space with a discrete time evolution consisting of
subsequent MC steps [thus our time unit is a MC Step (MCS)]. The
thermal averages become averages over the MC evolution after a
suitable thermalization of the system. We use a thermalization time 
constituted by $2000 \times n$ MCS (where $n$ is the total number of 
the spins of our system). Furthermore each point of the Markov chain must 
be generated after some MCS from the previous one so that each measure 
becomes uncorrelated from the others. Therefore between two measures we wait
a time longer than the autocorrelation time. In our specific
case this waiting time is chosen to be $500 \times n$ MCS. 
Analyzing the time fluctuations of the observables, we have verified that 
these choices of thermalization and autocorrelation time are optimal to have reliable measures. 

\begin{figure}[htbp]
\centering
\begin{tabular}{c}
 \epsfig{file=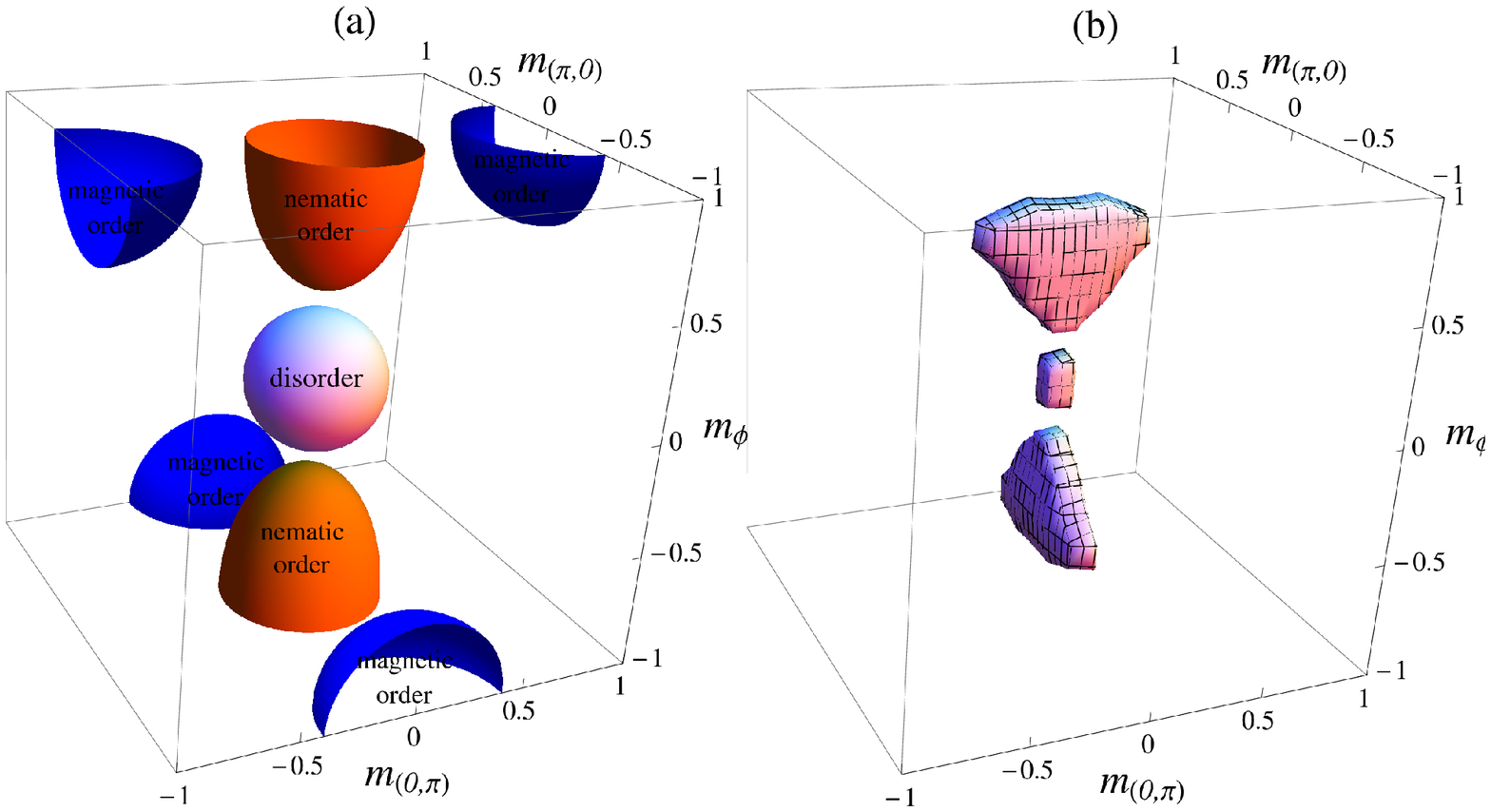,width=8.7cm} \\
 \epsfig{file=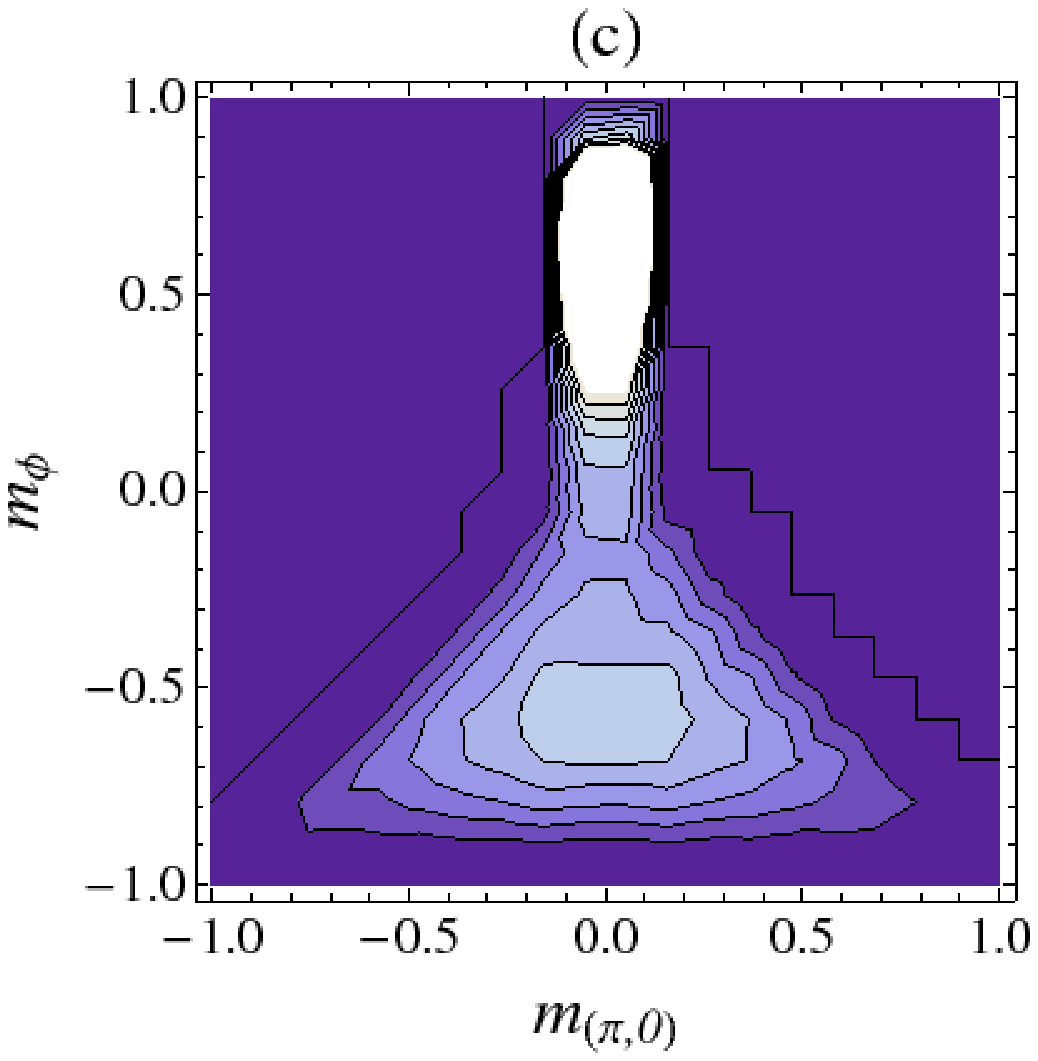,width=4.0cm}
\end{tabular}
\caption{(Color online) Phase space of the order parameters. (a) the spheres 
or ellipsoids schematically describe the different phases of the system. When the order parameter 
is distributed inside the central sphere the system is disordered. An order parameter 
inside the lightly shaded (orange online) 
ellipsoid indicates robust nematic  order, while when the order parameters are inside
the darker shaded (blue online) ellipsoids a magnetic state is realized. 
(b) Example of a MC simulation on a $50 \times 50$ lattice for $R=0.51$ 
and $k_BT/J_1= 0.52$. Statistically independent points where binned
according to the value of the 3D magnetization $\left( m_{(\p,0)},m_{(0,\p)},m_\f \right)$
in cells of size  $0.1 \times 0.1 \times 
0.1$. We show an isosurface with 500 points per bin. (c) Contour plot of the 
distribution projected onto the
$m_{(0,\p)}=0$ plane with bins of size $0.1 \times 0.1$. The bins with higher counts are the most 
bright.}
\label{fig:phases}
\end{figure}

Each configuration can be characterized by the instantaneous value of
the three magnetizations $ \left( m_{(\pi,0)},m_{(0,\pi)},m_\f
\right)$ defining a point in a three dimensional space. 
Fig.~\ref{fig:phases}(a) is a schematic representation of the regions where the points
corresponding to instantaneous configurations should be more dense in each one of the  phases 
summarized in Table \ref{tableI}.  

In order to characterize the different phases we compute the threedimensional (3D)
density distribution of points at each temperature of the simulation. 
This distribution can be visualized either by projection on one plane (Fig.~\ref{fig:phases}(c)) 
or by plotting 2D isosurface with a given number of points per unit
volume (Fig.~\ref{fig:phases}(b)). The points are expected to be distributed with a
Boltzmann weight $\exp\left[ -F(m_{(\pi,0)},m_{(0,\pi)},m_\f)/k_BT
\right]$  thus higher density of points corresponds to the lower free
energy $F$ and indicates the most stable phase. As it will be discussed 
in greater detail in the next section, the example 
in Fig.~\ref{fig:phases}(b,c)orresponds to the nematic phase.   
Since the distribution are rather flat for a given sign of the nematic
order parameter, the two-dimensional plots like the one of
Fig.~\ref{fig:phases}(c) allows to visualize the competition between the
nematic and the ordered phase.  Similar information can be cast in the form of a
scatter plot as in the example of Fig.~\ref{fig:scatterplot_no_nematic} below. 
Notice that the
competition between the nematic and the disordered phase are not well
characterized by this method.  In any case we have
checked all our results with the more rigorous 3D visualization
method. Specific details for each phase are discussed in next Section.

Due to the symmetries in the Hamiltonian for a generic point 
$ \left( m_{(\pi,0)},m_{(0,\pi)},m_\f \right)$
there are 7 other points which correspond to distinct configurations with the same energy. 
The symmetries are the reflection around the  $m_{(\pi,0)}=0$ and
$m_{(0,\pi)}=0$ planes and the reflection with respect to the $m_\f =0$ plane
followed by a $90^\circ$ rotation respect to the $m_\f$ axis. This
last rotation is due to the fact that the sign of $m_\f$ is linked to
the direction of the stripes. 
%In the $ \left[ m_{(\pi,0)},m_{(0,\pi)} \right]$ subspace of the 
%$ \left[ m_{(\pi,0)},m_{(0,\pi)},m_\f \right]$ space there is a reflection
%symmetry with respect to the two axis and an inversion symmetry with respect to the origin. 
%There is also a reflection symmetry with respect to the two bisectors if we map 
%$m_\f \ra -m_\f$. 
With these symmetries for each point obtained in 
the simulations we obtain 7 other symmetry-related points 
increasing the statistic and producing perfectly symmetric
distributions as expected for non symmetrized simulations performed for very long times. 

%The region with the highest
%distribution of points represents the equilibrium configuration, since
%it naturally minimizes the Helmholtz free energy.   

% If we plot the MC values of the three order parameters in 
%the three-dimensional $\left[ m_{(\p,0)},m_{(0,\p)},m_\f \right]$ space, we obtain a sort 
%of phase space exploration plot (shown in Fig.\ref{fig:phases}), each point of which 
%indicates the order of the system in a certain MCS  at a fixed temperature. 

\section{Results}

\begin{figure}[htbp]
\centering
\epsfig{file=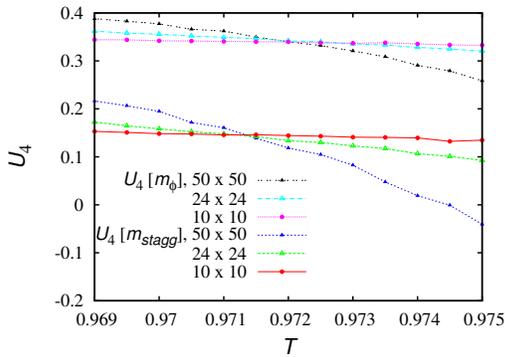,width=7cm}
\caption{(Color online) Temperature dependence of $U_4 \left[ m_{\f} \right]$ and 
$U_4 \left[ m_{stagg} \right]$, calculated for different  ($50 \times 50$, 
$24 \times 24$, $10 \times 10$) system sizes. 
%The temperature step is $5 \cdot 10^{-4}$.
}
\label{fig:BC}   
\end{figure}

To characterize the transitions to the ordered states, we determine the 
critical indices of the magnetic and nematic transitions using the Finite 
Size Scaling technique\cite{Newman1999Monte,Binder2002Monte}. Using our MC 
simulation, we calculate the temperature dependence of the interesting 
physical quantities of the system for three different lattice sizes: $10 \times 10$, 
$24 \times 24$, $50 \times 50$, using a value of $R$ not too close to
the $T=0$ critical  
point $R = 0.50$. Thus we choose $R = 0.60$. Although for this value of $R$, 
the magnetic transition is of first-order\cite{Lopez1993Firstorder}, 
since the transition is `weakly' of the first-order, we can still compute 
``critical indexes'' 
describing the substantial growth of the various susceptibilities near the
transition. We first calculate the temperature and size dependence  
of the Binder Cumulant\cite{Binder2002Monte}
\ie
U_4 \left[ m \right] = 1 - \frac{\lan m^4 \ran}{3 \, {\lan m^2 \ran}^2},
\fe
where $\langle m \rangle$ is the thermal average of the parameter we are interesting in (in our case the staggered 
magnetizations or the nematic one). The temperature at which the value of $U_4$ 
is the same for the three different lattice size defines the critical temperature 
of the transition. We show in Fig.\ref{fig:BC} the temperature dependence of
$U_4$ calculated for 
$\langle m_{\f} \rangle$, and for $\langle m_{stagg} \rangle$, that is the mean value between 
$\langle m_{(\p,0)} \rangle $ and $\langle m_{(0,\p)} \rangle$. Our 
dimensionless temperature $T$ is defined in units
of $|J_1|$ and taking a unit Boltzmann constant $k_B=1$.

From the plots in Fig. \ref{fig:BC}, one sees that the critical temperature of the magnetic 
transition is $T_{AF} \simeq 0.9715 \pm 0.0005$, while the critical temperature 
of the nematic transition is $T_{nem} \simeq 0.9720 \pm 0.0005$. Therefore, 
within our accuracy, the two transitions occur at the same temperature.

To establish the universality class of the two transitions we apply the 
scaling analysis to the magnetic susceptibilities
\begin{subequations}
\ie\label{eqn:chi_pi_0}
\chi_{(\p,0)} = \frac{\b}{N} \left[ \lan \sum_{i,j} \s_i \s_j 
\,{(-1)}^{(x_j - x_i)} \ran - { \lan \sum_{i} \s_i 
\, {(-1)}^{x_i} \ran}^2 \right];
\fe
\ie\label{eqn:chi_0_pi}
\chi_{(0, \pi)} = \frac{\b}{N} \left[ \lan \sum_{i,j} \s_i \s_j 
\,{(-1)}^{(y_j - y_i)} \ran - { \lan \sum_{i} \s_i 
\, {(-1)}^{y_i} \ran}^2 \right],
\fe
\end{subequations}
and to the nematic susceptibility
\ie\label{eqn:chi_phi}
\chi_{\f} = \frac{\b}{N} \left( \lan \sum_{i,j} \f_i \f_j \ran - { \lan \sum_i \f_i \ran } ^2 \right).
\fe

We consider temperatures sufficiently close to the critical
temperature, so that we can ignore the second term in the r.h.s.
of the two equations, ({\it i.e.} the square of the 
order parameter). Indeed at the critical temperature this term vanishes like 
$t^{2\b}$ (where $t=\left(T-T_{AF}\right)/T_{AF}$) and is negligible with respect to 
the correlation term (the first term in the r.h.s.) that tends to diverge 
like $t^{-\g}$.

\begin{figure}[htbp]
\centering
\epsfig{file=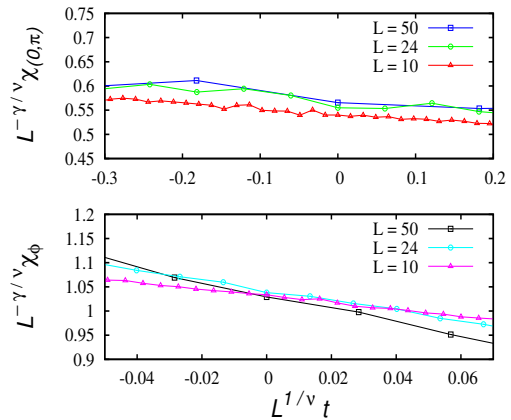,width=7.0cm,height=6.0cm}
\caption{(Color online) $\chi_{(0,\p)}$ and $\chi_\f$ scaling functions for 
lattice size $10 \times 10$ (triangles, red and magenta 
online), $24 \times 24$ (circles, green and ligth blue), $50 \times 50$
(squares, blue and black). The temperature step 
is $5 \cdot 10^{-4}$. We have fixed $\g = 7/6$, $\n = 2/3$, $T_{AF} = 0.9715$ for the 
magnetic transition, and $\g = 7.00/4.15$, $\n = 0.975$, $T_{nem} = 0.9720$ for the 
nematic transition.}
\label{fig:fss}   
\end{figure}

We report the scaling function $\tilde{\chi} = L^{-\g/\n} \chi(t)$ as a function 
of $L^{1/\n}t$ for the three different system sizes. 
The susceptibilities for systems of different size must coincide if the critical 
exponents $\g$, $\n$ and the critical temperature are exact
and the transition is of the second order. In Fig.\ref{fig:fss}
we show the scaling functions for the magnetic (we report only $\chi_{(0,\p)}$) 
and nematic suscetibilities.

We find that a
reasonable scaling is obtained with  $\g = 7/6$, $\n = 2/3$, $T_{AF}=
0.9715$. These critical exponents correspond to a  four-component Potts
model\cite{Wu1982Potts}.  Indeed, the magnetic order of our system is 
described by the  
two parameters $m_{(\p,0)}$ and $m_{(0,\p)}$, each parameter having
two possible values. Therefore the magnetic order 
parameter has 4 ``colors'' and threrefore one can naturally
 expect that the transition belongs
to the four-component Potts model universality class. 
We remark that the transition is weakly first-order\cite{Lopez1993Firstorder}
and therefore the scaling found can only be approximate. 
 
On the other hand it is natural to expect 
that the critical indexes of the nematic transition are those of the simple  
Ising model\cite{Huang1987Statistical}. We find $\g = 7.00/4.15$, $\n = 0.975$, 
$T_{nem} = 0.9720$, which are indeed rather close to the critical indexes of the 
twodimensional Ising model, but they do not quite coincide with them. Moreover 
the scaling functions of the three different system sizes do not collapse and 
are very close only when the reduced temperature $t$ is close to zero. 
Therefore the scaling found in this case is not ideal. This is
not surprising given that this transition is also weakly first-order (see below). 
Moreover the two transitions occur simultaneously and can influence
each other changing the universality class. In a more rigorous
approach the Ising and the Potts order parameters would form a single
order parameter with more components and a combined universality
class. However the scaling found above shows that the decomposition in 
Ising and Potts is a good approximation.

\begin{figure}[tbp]
\centering
\epsfig{file=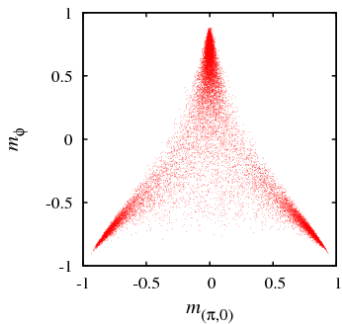,width=4.5cm} 
\caption{(Color online)
Scatter plot in the 2D-spaces 
$\left[  m_{(\p,0)},m_{\f} \right]$ resulting from the simulations of a 
$50 \times 50$ lattice with $R=0.60$, at the temperature $T = 0.9720$.}
\label{fig:scatterplot_no_nematic}   
\end{figure}

With the present choice of $R=0.60$ we find that there is no nematic
phase. For example  in 
Fig.\ref{fig:scatterplot_no_nematic} we show the scatter plot for the
system with size  
$L = 50$ at the temperature $T = 0.9720$, below which the 
nematic order parameter develops. 
There is no evidence of a nematic phase because there is no accumulation of points in 
the region with $m_{(\p,0)} \simeq 0$ and $m_{\phi}<0 $.  Notice that
the accumulation of points for  $m_{\phi}>0 $ corresponds to $(0,\p)$  magnetic
order. We find a qualitatively similar behavior at all intermediate
temperatures up to the disordered temperature. 

We now consider MC simulations for a $50 \times 50$ system with $R=0.51$, that is very close 
to the critical point at $R=0.50$. 
This case has already been shown as an example in Fig.\ref{fig:phases}. The
panel (b) shows a  
coexistence of disorder and nematic order at $T=0.52$, because we see a maximum in the 
disordered region and a maximum in the nematic region
($[m_{(\p,0)}\sim 0,\, m_\f \neq 0]$. 
Making other histograms with higher level surfaces we see that the
maximum of the disordered phase  
is lower than the one occurring in the nematic region. Therefore the
disordered phase is metastable.  
The occurrence of two maxima in the phase space population, that is two minima in the free energy,
also indicates that the transition from the disordered phase to the nematic phase is  
of the first-order.
%We have checked that by increasing the intensity level the central
%structure disappear before the of-center structures indicating that 
%the example shown corresponds to nematic order. The fact that two
%maxima coexist indicates that the transition is first-order. 
Fig.\ref{fig:phases}(c) is a contour plot of the counts for each bin in the 
$[m_{(\p,0)},m_\f]$ 2D-space. In principle we should also study the
$[m_{(0,\p)},m_\f]$ projection.   However, since
we made the two staggered magnetizations equivalent by exploiting the
symmetries of the problem,  we can focus on one of the two
projections only.  

\begin{figure}[htbp]
\centering
 \epsfig{file=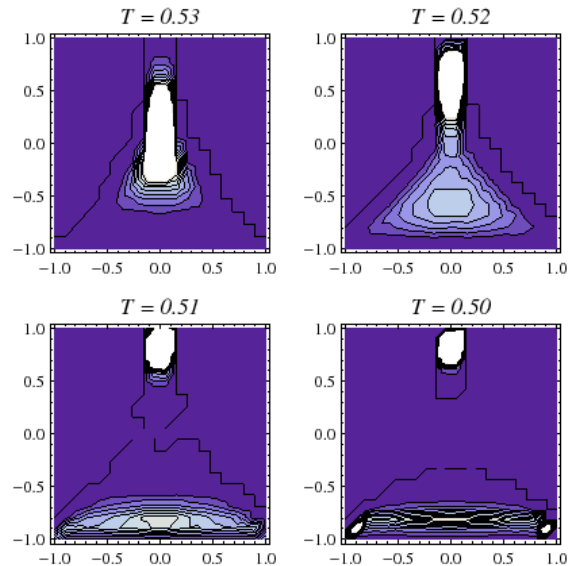,width=7.5cm}
\caption{(Color online) Level curves of the bin counts in the 2D-space $[m_{(\p,0)},m_\f]$ 
resulting from the simulations of a $50 \times 50$ lattice, with $R=0.51$. Are 
shown the results for $T=0.53$, $T=0.52$, $T=0.51$, $T=0.50$. The higher is the brightness of the bins and the higher is the 
values of the counts.  }
\label{fig:BinCounts_L50}
\end{figure}

Decreasing the temperature there is an increase of the counts ({\it
  i.e.}, configurations) 
related to the magnetic phase, that is identified by finite values of
$m_{(\p,0)}$. The temperature at which these counts become higher than
the ones at nematic region in the 3D representation,  correspond to
the critical temperature $T_{AF}$ of the nematic to magnetic
transition. 
We show in  Fig.\ref{fig:BinCounts_L50} the level curves of the bin counts in the $[m_{\p,0},m_\f]$ 2D-space, 
for a range of temperatures. 
Since also here there is coexistence of maxima in the
nematic and the magnetic region the transition  is  first-order.

\begin{figure}[htbp]
\centering
 \epsfig{file=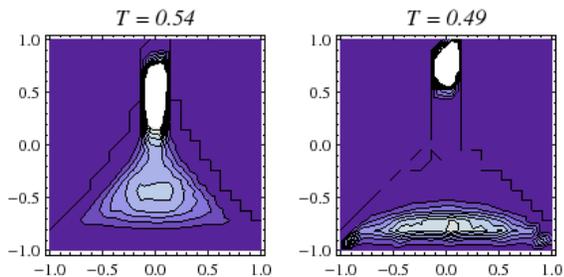,width=7.5cm}
\caption{ (Color online)
Level curves of the bin counts in the $[m_{(\p,0)},m_\f]$ 
2D-space resulting from the simulations of a $24 \times 24$ 
lattice, with $R=0.51$. The results for $T=0.54$ 
and $T=0.49$ are reported. The bins with highest counts are the most bright.}
\label{fig:BinCounts_L24}
\end{figure}

To compute the temperature range of the nematic phase we study the histograms of the simulations 
at different temperatures (we use $\D T = 0.01$ steps in temperature).     
A close inspection of the level surfaces of the histograms in 3d-space [like the one shown in 
Fig.\ref{fig:phases}(b)], we can state that at $T=0.53$ the maximum in correspondence 
to the disordered phase is higher than the one corresponding to the nematic phase. 
Similarly at $T=0.50$ the maximum of the smectic phase is the highest. Hence the range 
of temperatures in which the system has nematic order only includes the $T=0.51$ and 
$0.52$ temperatures.
\begin{figure}[tbp]
\centering
\epsfig{file=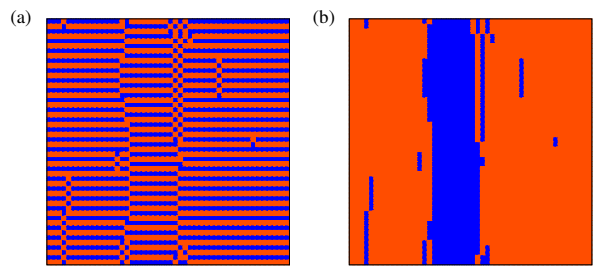,width=8cm}
\caption{(Color online) Magnetic configurations 
of a $50\times 50$ systems with $T=0.52$;
(a) real spin configuration, where the red squares represent the spin-up, 
while the blue squares the spin-down, (b) staggered configuration along $y$ axis, 
where the squares are red if $\s_i \cdot {(-1)}^{y_i} = +1$, while they are blue if 
$\s_i \cdot {(-1)}^{y_i} = -1$.}
\label{fig:config_ret}   
\end{figure}
We show in Fig.\ref{fig:config_ret} the magnetic configuration at the end of the 
simulation at the temperature $T=0.52$.  
One can clearly see that rotational symmetry is broken, but the system does not display
a complete magnetic order, because the average value of the staggered
magnetization,  which define the magnetic order is small.
We also studied  smaller systems with  $24 \times 24$ and 
$10 \times 10$ lattice sites. Performing the same analysis as in the previous case, 
we observe that the range of temperatures where nematic order occurs is larger than 
in the $50 \times 50$ system.
We report in Fig.\ref{fig:BinCounts_L24} the same plots as in Fig.\ref{fig:BinCounts_L50} 
for the  $24 \times 24$ system at the two temperatures delimiting the nematic range.
At the temperatures considered in Fig.\ref{fig:BinCounts_L24} there is still an evident 
nematic phase.
For a $10 \times 10$ lattice, the nematic range is 
even larger. We can therefore conclude that the smaller is the system size and the stronger 
is the tendency to form a nematic phase. 

To explain this result we compute the energy cost to form a domain 
like the one displayed in Fig.\ref{fig:goccia_nematica} on the perfect
$(\pi,0)$ order. Notice that this corresponds to a change of phase of
the magnetic order but without rotation of the direction of the stripes. 
These domains will be naturally elongated and 
we aim to find the typical aspect ratio.  

We consider for simplicity a  rectangular domain like the one shown
in Fig.\ref{fig:goccia_nematica}. 
It is easy to see that the boundary energy is anisotropic.
 With respect to the ground state (complete smectic phase),
a spin located along the vertical walls brings along an 
 increase of energy per unit length of
domain boundary $\D E_{ver} = 2J_1 + 4 J_2$.
On the other hand, the increase of energy per unit length due to a spin located 
along the horizontal wall is given by $\D E_{hor} = -2J_1 + 4 J_2$. 
We call $l_\perp$ 
the length of the walls perpendicular to the stripes, and $l_{\|}$ the
length of the walls parallel to the stripe. The number of spins along the horizontal 
walls of the rectangular domain will be $2l_\perp$, while along the vertical walls 
they are $2l_\|$. Thus the boundary energy of the domain is given by
\ie
E_{bnd} = 2l_\perp (-2 J_1 + 4 J_2) + 2l_\| (2 J_1 + 4 J_2).
\fe
We now determine what is the most stable configuration for this type
of domains.  
To this purpose we minimize $E_{bnd}$ varying its aspect ratio
$l_\perp/l_\|$, while maintaining its area $l_\perp l_\|$  
fixed. We find,
\ie\label{eq:l_ratio}
\frac{l_\perp}{l_\|} = \frac{2 J_2 + J_1}{2 J_2 - J_1} = \frac{2R+1}{2R-1}.
\fe
Although this computation applies to the ordered phase, as the
temperature is lowered from the disordered phase, we expect that 
elongated domains are formed with approximately the same aspect
ratio. This defines two typical length scales $\c_\perp$ and 
$\c_\|$ corresponding to two different correlation lengths satisfying 
\ie\label{eq:csi_ratio}
\frac{\c_\perp}{\c_\|} = \frac{l_\perp}{l_\|}.
\fe
Approaching the N\'eel temperature $T_{AF}$ from above,
the correlation lengths should be described by 
\begin{subequations}
\ie
\c_\perp = \frac{\c_\perp^0}{{|T-T_{AF}|}^\n}; 
\fe
\ie
\c_\| = \frac{\c_\|^0}{{|T-T_{AF}|}^\n}, 
\fe
\end{subequations}
where $\c_\perp^0$ and $\c_\|^0$ are simple proportionality factors,
whose ratio is expected to be given by  
Eqs.(\ref{eq:l_ratio}) and (\ref{eq:csi_ratio}). 

For values of $R$ near $0.5$, we have  $\c_\perp^0 \gg \c_\|^0$ 
and there will be a range of temperatures for which the linear system
size, $L$, satisfies  
\begin{equation}
  \label{eq:nematic}
\c_\| < L < \c_\perp.   
\end{equation}
This means that 
the system is ordered in the direction perpendicular to the stripes 
but it is still disordered along the direction of the stripes.
This is precisely the nematic phase. Indeed we can see in the snapshot
of Fig.~\ref{fig:config_ret} (b) that the domains are elongated and span
one linear dimension of the sample, while the system has a correlation
smaller than $L$ in the  direction parallel to the stripes.

As the system size increases the temperature range in which
Eq.~(\ref{eq:nematic}) is satisfied decreases, i.e. the nematic
temperature window decreases with system size, in agreement
with our simulations. 

Simple arguments show that the value of $R$ in FeAs planes is actually close 
to its critical value $R = 0.5$\cite{Dai}.
These results can also shed light on the experiments 
by Jesche {\it et al.}\cite{Jesche2010Coupling} and Liang {\it et
  al.}. The former found 
that the splitting between the structural 
and the magnetic transition 
 decreases as the sample quality is 
increased while the latter finds that the anisotropy in transport
decreases.  It is quite natural to assume that 
grain boundaries, dislocations and other defects
will disrupt the perfect crystal order and introduce an extrinsic scale $L$
roughly given by the average distance among defects. As the sample is
cooled from the disordered phase, at some point the longitudinal 
correlation length will exceed the scale $L$. Then at this scale $L$
the $C_4$ symmetry of the crystal will be spontaneously broken and
the system will undergo a spontaneous orthorhombic distortion possibly 
with twining or domain formation at the scale $L$. In the
perpendicular direction the system will remain essentially disordered
preventing the appearance of an elastic signal in magnetic neutron
scattering. At a lower temperature the system will become magnetically 
ordered in both directions. 

This picture applies for weak magnetoelastic coupling. 
As we will show in the next Section, when the
the spin system is strongly coupled to the lattice the structural 
and nematic transitions occur at the same temperature and 
the splitting between the structural and the magnetic transition 
coincides with the splitting between the nematic and the magnetic transition. 

\section{Spin-lattice coupling}

In this Section we consider lattice deformations and their
effects on the spin system.  In the continuum the distortions can be
characterized by the symmetric strain tensor 
\ie
u_{\mu\nu} = \frac{1}{2} \left( \frac{\partial u_\mu}{\partial x_\nu} 
+ \frac{\partial u_\nu}{\partial x_\mu} \right),
\fe
where $\mu$, $\nu$ label the Cartesian directions.

In a 2D system it is convenient to work with the following strains, 
\ie\label{eq:deformations} 
e_1 = u_{xx}+u_{yy}, 
\quad e_2 = u_{xx} - u_{yy}, \quad e_3 =\sqrt{2} u_{xy}, 
\fe 
which describe dilational, deviatoric and shear deformations respectively.  

To include magnetoelastic effects  
in our frustrated Ising model, we assume that the coupling 
constant $J_{1i,\nu}$ among atom $i$ and the nearest-neighbour atom in
the $\nu$ direction  depends linearly on the deformation according to
\begin{subequations}\label{eq:coupling_development}
\ie
J_{1i,x}=J_1 + 2 \a u_{xx}  = J_1 + \a (e_1+e_2), 
\fe
\ie
J_{1i,y} = J_1 + 2 \a u_{yy} = J_1 + \a (e_1-e_2). 
\fe
\end{subequations}

For simplicity we will restrict to uniform strains which is enough to
describe the observed structural transition. At lowest order the shear
strain does not couple with the magnetism. The dilation strain $e_1$
simply adds to $J_1$ and describes symmetry preserving changes in the
volume which are interesting but not the focus of the present
work. Therefore we set $e_1=e_3=0$ and consider only $\epsilon\equiv e_2$.
The latter is the order parameter of the observed  structural
transition\cite{Krellner2008Magnetic,delaCruz2008Magnetic} which for a
FeAs layers implies a deformation from a square lattice to a
rectangular lattice.   Within this approximation we obtain the
magnetoelastic term 
\ie
\label{eq:spin-latt-coupl}
H_{M-el} = - \a \e \sum_{i=1}^N \s_i (\s_{i+x} - \s_{i+y}) 
= -2 \a \e \sum_{i=1}^N \phi_i.
\fe 
In addition we must consider the  elastic energy which is given by 
\ie\label{eqn:H_el}
H_{el}=\frac{1}{2}N \mu_0 {\e}^2,
\fe
where $\mu_0$ is a shear modulus at $T=0$.

We thus obtain a sum of 
three terms: the purely magnetic term
Eq.~(\ref{eq:H_frustrates_ising}), a purely elastic term
Eq.~(\ref{eqn:H_el}) and the magnetoelastic term 
Eq.~(\ref{eq:spin-latt-coupl}).
This last term contributes to the internal energy $U \equiv 
\langle H \rangle$, by an amount $\langle H_{M-el} 
\rangle = -2 N \a \e \langle m_\f \rangle $, directly involving the thermal 
average of the nematic magnetization $\langle m_\f \rangle$. 
Thus the nematic and structural order
parameter couple linearly with each other.  This 
implies that the nematic critical temperature should
coincide with the critical temperature of the structural transition. 

To identify the stable equilibrium states of the system at finite 
temperature, we now study the Helmholtz free energy. 
We start from values of temperature higher than both the 
critical temperature $T_{AF}$ of the magnetic transition and 
the critical temperature $T_{S}$ of the structural transition. 
Therefore no magnetic order and no structural deformations are present
and the free energy will be minimum for 
$\langle m_\f \rangle=0$ and $\e = 0$. If we consider small variations of the two 
parameters $\langle m_\f \rangle$ and $\e$ around their zero values, we can expand 
the free energy up to second order in the two variables thereby obtaining
\ie\label{eq:F_development}
\frac{\delta F(\langle m_\f \rangle, \e)}N = \frac{1}{2}   (\chi_\f^0)^{-1} 
{\langle m_\f \rangle}^2 +  
 2\alpha \langle m_\f \rangle \, \e + \frac{1}{2}  \m \, {\e}^2,
\fe   

Adding external fields coupling linearly with the order parameters one
can see that $\chi_\f^0$ is the same  nematic susceptibility 
appearing in Eq. (\ref{eqn:chi_phi}).
The last coefficient $\m$ is the shear modulus at finite 
temperature  and in the absence of magnetoelastic coupling.
Within our approximations we can consider this term equal to the
coefficient $\m_0$ of  
Eq.(\ref{eqn:H_el}), since the elastic term (\ref{eqn:H_el}) of 
the Hamiltonian coincides with the elastic free energy 
$F_{el} = \frac{1}{2} N \m {\e}^2$. Indeed, since
the deformation $\e$ is constant along the whole system, the lattice 
entropy is a quantity of order 1 
and can safely be neglected when extensive quantities of order $N$
are considered. 
Minimizing Eq.~(\ref{eq:F_development}) with respect to the nematic 
magnetization $\langle m_\f \rangle$, we obtain  
\ie
\langle m_\f \rangle = -2 \a \, \e \, \chi_\f^0.
\fe   
Replacing this expression of $\langle m_\f \rangle$ into 
Eq.(\ref{eq:F_development}) we find
\ie\label{eq:F_development_c}
\delta F(\e) = \frac{1}{2} N \left( \m_0 - 4 \a^2\chi_\f^0 \right) \, 
{\e}^2 =  \frac{1}{2} N \, \m_{tot} \, {\e}^2,
\fe
where $\m_{tot}(T)$ is the temperature dependent shear modulus, which takes
into account the renormalization due to magnetoelastic coupling.  

Introducing an effective coupling constant $\l = 4 \a^2 / \m_0$, 
from Eq.~(\ref{eq:F_development_c}) one obtains a generalized Stoner
criterium. If $\chi_\f^0(T) < 1 / \l$, the system is stable, whereas,
if $\chi_\f^0(T) > 1 / \l$, the system becomes unstable and 
gains energy by making a deformation $\e$. The temperature at 
which $\chi_\f^0(T) = 1 / \l$ defines the critical temperature 
$T_S^{2nd}$ of the (second-order) structural transition.

Let's assume that in the absence of magnetoelastic coupling, the 
transitions are second order and that the nematic and magnetic
susceptibility diverge at the same temperature, i.e. that there is no
true thermodynamic
nematic phase for $\alpha=0$. The above argument  shows that for small
magnetoelastic coupling the nematic and the orthorhombic  orders
occur  simultaneously at the temperature $T_S^{2nd}$ 
higher than that of magnetic order.  
Indeed in the disordered phase, the nematic susceptibility $\chi_\f^0$ will grow 
upon lowering the temperature and the instability (Stoner-like)
condition will be 
satisfied while the magnetic susceptibility is still finite. 
Although the presence of the structural distortion will tend to increase the Neel
temperature $T_{AF}$, this increased temperature can never reach
$T_S^{2nd}$ since $\epsilon$ vanishes at $T_S^{2nd}$. Therefore a finite
temperature window should appear between  $T_S^{2nd}$ and $T_{AF}$. 
Thus  the magnetoelastic coupling can induce a nematic
phase. Of course, if the nematic phase were already existing for $\alpha=0$, a
small magnetoelastic coupling would tend to widen the nematic
temperature window.

Notice that the above arguments are general and do not depend
on the details of the model (e.g.  Ising vs. Heisenberg) but on the
second order character of the transition. We will show below that the weakly
first-order character is enough to change this scenario. For a first
order transition occurring at $T_{S}^{1st}$ the temperature 
$T_{S}^{2nd}$ where the shear modulus vanishes  satisfy $T_{S}^{2nd}<T_{S}^{1st}$ 
and corresponds to the limit of
stability of the metastable disordered phase. 

In order to investigate the feedback effect of the lattice
on the magnetic transition and the effect of the first-order
character we carry out a Monte Carlo study of the magnetoelastically coupled 
Hamiltonian. 
To take into account the thermal fluctuation of the deformation parameter 
$\D J = \a \e$, every $n$ Monte Carlo step (where $n$ is the number 
of the lattice sites) we allow for random variations of this parameter in 
a certain range (we use $[-0.2:0.2]$). Each variation of $\D J$ 
is accepted or refused according to Metropolis algorithm with 
the energy given by Eqs.(\ref{eq:H_frustrates_ising}), (\ref{eq:spin-latt-coupl}) and (\ref{eqn:H_el}). 
We perform temperature runs (with a step $\D T = 0.01$) of a $50 \times 50$ 
system with $R = 0.51$, for several values of $\l$. 

\begin{figure}[htbp]
\centering
\epsfig{file=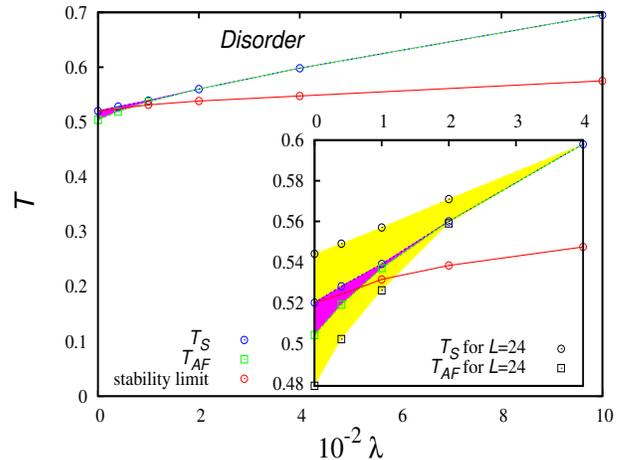,width=8.5cm,height=6.5cm}
\caption{(Color online) Phase diagram of the system with $R=0.51$ in the 
$\l$ {\it vs.} $T]$ space. We report the numerical results for the 
structural (or nematic) critical temperatures (open circles, blue online) 
and the magnetic critical temperatures (open squares, green online), and 
the structural critical temperatures in the stability limit (open circles,
red online, joined by the solid line).
The shaded (magenta online) region represents the nematic range 
in a $50 \times 50$ lattice. The inset displays an expansion of the 
nematic region of the phase diagram. The nematic 
range in a $24 \times 24$ system (darker shaded region, yellow online) 
is also shown.}
\label{fig:T_vs_lambda_phase_diagram}   
\end{figure}

The phase diagram 
in the $[\l,T]$ space, resulting from our simulations, is 
reported in Fig.\ref{fig:T_vs_lambda_phase_diagram}.  
We see that increasing $\l$, the temperature  interval with nematic order
shrinks and it disappears completely for $\l = 2.0 \cdot 10^{-2}$. Thus 
the largest nematic window is for $\l = 0$ and it correspond to the case 
studied  in the previous section. This goes against our original
expectation to find an enhanced nematic phase in the presence of the
magnetoelastic coupling. Such a failure is clearly due to the first
order character of the transition which gets enhanced by the
magnetoelastic coupling. This can be recognized from the fact that 
 $T_{S}^{2nd}$ and $T_{S}^{1st}$ practically coincide for $\l = 0$,
 while $T_{S}^{1st}$ becomes substantially larger than
 $T_{S}^{2nd}$ as $\lambda$ increases. This is also supported by the 
presence of thermal hysteresis of the order parameters 
that we observe at high values of $\l$.

\begin{figure}[htbp]
\centering
\epsfig{file=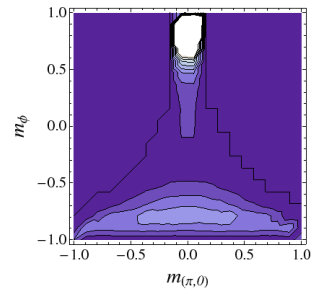,width=4cm}
\caption{(Color online) Contour plot of the order parameter distribution
    projected in the $[m_{(\p,0)},m_\f]$ plane resulting from the
    simulations of a $50 \times 50$ lattice, with $R=0.51$ and $\l =
    1.0 \cdot 10^{-2}$, at $T=0.538$.} 
\label{fig:bin_counts_spin-lattice}   
\end{figure}

Notice that as expected 
$T_{S}^{2nd}$ increases with $\lambda$ but $T_{AF}$ increases more and
due to the first-order character of the structural transition this
renormalization of $T_{AF}$ can reach $T_{S}^{1st}$ closing the nematic window. 
As discussed in the previous Section, this window is larger for smaller
system sizes. Indeed, as it is shown in the inset of
Fig.\ref{fig:T_vs_lambda_phase_diagram}), in
a $24 \times 24$ system  a larger nematic region 
in the $[\l,T]$ space is  found.
Although increasing $\lambda$ leads to the closing of the nematic
window, a nematic phase does exist at small spin-lattice coupling:
We report in Fig.\ref{fig:bin_counts_spin-lattice} a contour plot of 
the order parameter distribution function for  $\l = 1.0 \cdot
10^{-2}$ which shows clearly that the system is in a nematic phase. 

In short we can state that for small $\l$, transitions 
from disordered phase to nematic phase and from nematic phase 
to magnetic phase are both only `weakly' of the first-order. Therefore
our schematic model system grossly reproduces the features of the 1111
\cite{delaCruz2008Magnetic} or of Ba(Fe$_{1-x}$Co$_x$)$_2$As$_2$ 
\cite{fernandes1,nandi} materials, where a separation between a
 second-order structural transition and a  second-order magnetic transition
is  observed. In particular we notice that, upon increasing doping
in Ba(Fe$_{1-x}$Co$_x$)$_2$As$_2$ the nematic window increases becoming
largest at $x\gtrsim 0.06$, near optimal doping \cite{nandi}. This strengthening
of the nematic state finds a rational within our findings because
increasing doping naturally leads to a decrease 
of magnetoelastic couping $\lambda$ and to an increase of disorder.
As found in our schematic model, both these effects increase the 
nematic region between the structural and the magnetic transitions
(cf. the inset of Fig. \ref{fig:T_vs_lambda_phase_diagram}).
We also notice that estimates from shear modulus measurements in
Ba(Fe$_{1-x}$Co$_x$)$_2$As$_2$ give typical values of $\lambda\lesssim
5\cdot 10^{-3}$, substantially smaller than the values at which
the nematic windows closes in our Fig. \ref{fig:T_vs_lambda_phase_diagram}). 

When instead we choose a high value of $\l$, we obtain a 
system which grossly reproduces the behavior of many 122 
materials\cite{Krellner2008Magnetic} where the structural and 
magnetic transitions are strongly first-order and occur 
at the same temperature.  
We furthermore notice that when the transitions become strongly 
of the first-order, the value of the critical temperature 
increases. Although the precise value of the critical temperature will
depend on many other parameters it is interesting that  the critical
temperature of the `122' materials are substantially 
higher than the critical temperatures of the `1111' materials.

\section{Conclusions}

In this paper we investigated numerically and with simple analytic 
arguments the possible occurrence of a
nematic phase and the associated splitting of the structural and
magnetic transitions in FeAs planes. We analyzed 
a schematic model of Ising spins with nearest and next-nearest
neighbor couplings. The effects 
of a spin-lattice coupling has also been considered. 
First of all we find that for $R$ larger than (but close to) 
1/2 the model displays a
weakly first-order transition. The very weak first-order character
found in the absence of lattice coupling  
allows to study critical exponents which for the magnetic order
parameter, are close to the ones of a  
4-state Potts model in agreement with simple symmetry considerations.

One can then remark that the magnetic Potts order parameter 
can order at finite temperature even in two dimensions and therefore
long range magnetic order  
effectively competes with the Ising nematic order. This model,
therefore, has the 
double advantage of being more easily solved in numerical MC
approaches and of being  an ``acid'' test for the occurrence of
nematic order. In addition, as we argue in the introduction, 
collective magnetism produces a system with large connectivity  
of the magnetic Hamiltonian, which makes the thermal transition rather
mean-field like.  
We argue that such mean-field transition in a real lattice with 3D
couplings and weak anisotropies is more similar to a 2D Ising 
transition than to a
2D Heisenberg transition. Thus, despite the apparent obvious
difference in symmetry between our model and a real system, the model
for some aspects is expected to be more realistic than a frustrated 2D
Heisenberg model. 

Our numerical study indicates that in the region of the stripe ground
state, but close to the transition to the N\'eel AF phase ($R\gtrsim 0.5$),
nematic states can form due to finite size effects in some temperature range
above the magnetic order (the smaller is the length scale
and the larger is the temperature range). We argue that such finite
size effects are a proxy for the effect of crystal imperfections
which disrupt the perfect lattice order and provide a natural
explanation for the results of Jesche {\it et
  al.}\cite{Jesche2010Coupling} and  Liang {\it et
  al.}\cite{liang} who find more robust nematic signatures in 
the worst quality samples. 

The increase in resistivity anisotropy\cite{liang} with decreasing
sample quality has also been explained by  Fernandes {\it et
  al.}\cite{fernandes2} in terms of the interplay between
scattering of carriers by spin fluctuations and
disorder.\cite{fernandes2} We believe both
Ref.~\onlinecite{fernandes2} mechanism and the
present one can contribute to the
observed effect. More experimental and theoretical work would be needed
to determine which one will be dominant in the different regimes.

We have neglected all Heisenberg physics which is expected to enhance
the tendency to nematic phases.\cite{larkin,Fang2008Theory} However our
arguments for the enhancement of the nematic phase in imperfect
systems also applies to other models. For example also   
in a more realistic 3D Heisenberg model, possibly with long range
couplings and anisotropies, the domains will be elongated and there
will be two characteristic length scales, so that our arguments remain
valid. 

It is sometimes argued that, since the structural transition occurs at
a higher temperature than the magnetic transition, it is the structure
that  drives the magnetism. This argument is clearly wrong and our
model is another counterexample which show that the structural
transition can occur at a 
higher temperature than the magnetism although it is driven by the
magnetism itself. This is because the structural transition is driven by the
cohesive energy which is determined by short range correlations like
$\langle \sigma_i \sigma_j \rangle$ with $i$ and $j$ close
neighbors. Thus cooling the system one can have a robust short range $\langle \sigma_i \sigma_j
\rangle$, which renders the lattice unstable, before the system has
long range magnetic order. Formalizing this argument in the case of
second order transitions, 
one obtains the additional result  that the coupling to the lattice
favors the nematic state. However in the simulations we find the opposite
because the lattice tends to make the transition more first-order like. This
suggest that the nematic state observed is mainly driven by the
electronic degrees of freedom and that the lattice act as a
spectator, as suggested in Ref.~\onlinecite{Fang2008Theory}.
 
The different behavior found for weak and strong magnetoelastic
coupling corresponds well with the behavior found on different pnictide
families.  Indeed 122 materials have 
 larger critical temperature, first-order transitions a non-nematic
 state, as found for  strong magnetoelastic
coupling. On the other hand  1111 systems have 
lower transition temperatures, second-order or weakly first-order
transitions and a nematic state as found for weak  magnetoelastic
coupling. 

\acknowledgments We thank  Andrea Pelissetto and Claudio
Castellani for useful discussions. J.L. thanks Matthias Vojta for
useful insights. M.G and J.L. acknowledge financial support form the
 MIUR-PRIN07 prot. $2007FW3MJX_003$.

\bibliographystyle{prsty_no_etal}
% \bibliography{matt84}

\end{document}